\documentclass[twocolumn,showpacs,amsmath,amssymb,superscriptaddress]{revtex4-1}
\usepackage{xcolor}
\definecolor{maroon}{RGB}{150,0,0}
\definecolor{dblue}{RGB}{0,0,150}
\usepackage[hyperindex,breaklinks,colorlinks=true,linkcolor=dblue,citecolor=maroon,urlcolor=maroon]{hyperref}
\usepackage{graphicx}
\usepackage{amscd,amsfonts,color,bbm, amssymb, amsthm}
\newcommand{\bra}[1]{{\left\langle #1 \right|}}
\newcommand{\ket}[1]{{\left| #1 \right\rangle}}
\newcommand{\T}{\mbox{$\mathrm{tr}$}}
\def\hcal{{\cal H}}

\def\>{\rangle}
\def\<{\langle}
\def\tr{{\rm Tr}}

\def\ot{\otimes}
\newcommand{\etal}{{\it et~al.~}}

\begin{document}
\title{Increasing distillable key rate from bound entangled
states by using local filtration}
\author{Mayank Mishra}
\email{mayank1@email.arizona.edu}
\altaffiliation{present address: College of Optical
Sciences, University of Arizona,
1630 East University Boulevard, Tucson, AZ 85721.}
\affiliation{Department of Physical Sciences, Indian
Institute of Science Education and Research Mohali, Mohali
140306, India}
\author{Ritabrata Sengupta}
\affiliation{Department of Mathematical Sciences, Indian
Institute of
Science Education \& Research (IISER) Berhampur, Transit
campus, Govt. ITI, NH 59, Engineering school junction,
Berhampur 760 010, Odisha, India}
\email{rb@iiserbpr.ac.in}
\author{Arvind}
\email{arvind@iisermohali.ac.in}
\affiliation{Department of Physical Sciences, Indian
Institute of Science Education and Research Mohali, Mohali
140306, India}
\date{\today}
\begin{abstract}
We show the enhancement of distillable key rate for quantum
key distribution(QKD), by local filtering, for several bound
entangled states. Through our work it becomes evident that
the local filtration operations, while transforming one
bound entangled state to another, have the potential to
increase the utility of the new state for QKD. We
demonstrate three examples of `one way distillable key rate'
enhancement by local filtering and in this process, discover
new bound entangled states which are key distillable.
\end{abstract}
\maketitle
\section{Introduction}
\label{intr0}
A perfect cryptography protocol can be set up if one can
distribute a private key between the trusted parties
interested in secure communication~\cite{MR1302169}.  It has
been established that only quantum key distribution (QKD)
protocols are fundamentally secure, as they draw their
security from the laws of quantum physics, as opposed to
their classical counterparts  where the security is based on
the impossibility of solving certain mathematical problems
in polynomial time~\cite{NC,EK,DEJCPS}.  The QKD protocols
are either of the prepare and measure type, such as the BB84
protocol~\cite{BB84, BB84-1} or the entanglement assisted
protocols such as the E91 protocol~\cite{EK}. These two
classes of protocols are intimately connected and
entanglement is considered  a fundamental resource for
QKD~\cite{review1}.

Bipartite quantum entanglement involves non-classical
correlations between two parties and can occur in pure as
well mixed states~\cite{epr,schrodinger,RevModPhys.81.865}.
A powerful tool to identify entanglement is the transpose
operation, where entangled quantum states can transform to
non-states, when we apply the transpose operation on one of
the sub systems~\cite{P1}. The states whose density
operators become negative under such partial transposition
are entangled and are called negative under partial
transpose (NPT) and those which remain positive and valid
are called positive under partial transpose (PPT).  From the
work of Peres and Horodeckis', it became clear that partial
transpose is necessary and sufficient only for $2\otimes 2$
or $2\otimes 3$ bipartite systems, while for higher
dimensional systems there can be entangled states which are
PPT~\cite{H1}. Due to the existence of PPT entangled states,
the geometry of states in composite dimensions other than
$4$ and $6$ has not been fully understood. The examples and
methods to construct classes of PPT entangled states are
sparse~\cite{arvind-r,saro_rb1,saro_rb2}, although  a
plethora of literature exists on the
subject~\cite{RevModPhys.81.865,guth,darek-sar}.

Ideally, quantum information tasks require pure entangled
states. Since mixedness caused by the environment is
unavoidable, for operational purposes it is necessary to
recreate pure copies of the required entangled state via
distillation protocols~\cite{dist-1,dist-2}. Entangled
states from which we cannot distill any pure entangled
states are called bound entangled and their entanglement is
called bound entanglement.  All pure entangled states are
NPT, and distillable entangled states from which maximally
entangled states can be distilled, are also NPT.  It can be
shown that pure entangled states can not be distilled from
PPT entangled states and hence PPT entangled states are
bound entangled~\cite{dist-3}.

Contrary to expectation, bound entanglement has been
found to be a useful resource for performing a  number of
quantum information processing tasks.  It was shown by Wild
and Hsieh~\cite{Wilde_2010} that bound entanglement can be
used for establishing super-activation of quantum channels
by using the Smith-Ward protocol~\cite{Smith1812}. Using
tensor products of bound entangled states to obtain
distillable states and aspects related to super-activation
and super-additivity are discussed
in~\cite{PhysRevLett.90.107901, PhysRevLett.108.190501}.
Usefulness of bound entanglement in metrology has been
observed by Czekaj et.  al.~\cite{PhysRevA.92.062303} where
they have given an example of a family of bound entangled
states which can be used in quantum enhanced metrology with
the precision advantage approaching the Heisenberg limit.
Further, T\'oth and V\'ertesi~\cite{PhysRevLett.120.020506}
demonstrated that multipartite quantum states that have a
positive partial transpose with respect to all
bi-partitions, can outperform separable states in linear
interferometers.  Bound entanglement has also been shown to
be a useful resource for quantum heat
engines~\cite{MR4031322}.

For QKD protocols, for a long time it was
believed that distillable entanglement from which one can
obtain maximally entangled states is an essential 
resource~\cite{EK} and bound entanglement is not
such a resource. Thus it came as a surprise when it was
shown that PPT bound entangled states may also be useful for
QKD~\cite{HPHH, HPHH05, ozols}. 
A key role is played by a new class of states called private
states~\cite{HPHH,HPHH05,HHHO05a} which can be used to carry
out QKD. These states need not be maximally entangled  but
still can be used to obtain a perfectly correlated key that
is completely uncorrelated to any eavesdropper~\cite{HPHH,
HPHH05}. Further, it was demonstrated that one can find PPT
bound entangled states that are arbitrarily close in trace
norm to the private states~\cite{ozols}. Thus using the idea
of private states it was shown that the PPT bound entangled
states can also be used for QKD~\cite{HPHH, HPHH05}.  The
initial class of key distillable bound entangled states
presented in~\cite{HPHH,HHHO05a} were of large dimensions and
hence of limited use for experimental applications. A new
class of low dimensional bound entangled states which can be
used for QKD key was introduced in~\citep{DJJTS07}. These
have been further studied
in~\cite{ozols,PhysRevA.85.012330,
MR2607919}.

Evaluating a PPT entangled states  for its utility for QKD is not
always straightforward and there are more than one criteria
being used in the literature. The two most important
parameters in this context are `distillable key rate'
($K_D$) defined in reference~\cite{HPHH} and  the `one way
distillable key rate' ($K_{D}^{DW}$) given in the
Devetak-Winter protocol~\citep{DW}. As per the definition given
in~\cite{HPHH}, a state is declared useful for QKD if
$K_{D}>0$. However for the states defined
in~\citep{DJJTS07}, while it has been shown that $K_{D}>0$,
we have $K_{D}^{DW}<0$~\cite{DW}.  The situation
therefore is delicate and different definitions may not
always agree~\cite{ozols}. However, for the cases that we
consider in our work  we have $K_D > K_D^{DW}$ and thus we
use the `one way distillable key rate' as a lower 
bound of the available key from a PPT state.

Local filtration is a process where starting with an
ensemble of bipartite states, the members are  selected or
discarded based on the results of certain local
measurements, thereby obtaining a new smaller (filtered)
ensemble of bipartite states~\cite{verfil}.  While local
filtration can not change the nature of entanglement (PPT or
NPT), the filtered ensemble can have different properties as
compared to the original one.
Particularly, such a filtration has been used to enhance the
Bell violation~\cite{gsain}, for detecting the
entanglement of PPT entangled states~\cite{Das2017}
and for increasing the usefulness of 
quantum states for QKD~\cite{sandeep_qkd}.  In this paper we use the local
filtration on different PPT entangled states and obtain
transformed states that have higher one way distillable key
rate. We demonstrate our method  by considering three
concrete examples of PPT entangled states, some of which are
available in the literature. First, we apply local
filtration to a family of  states closely related to the
states given in~\citep{DJJTS07} and show that under certain
conditions the one way distillable key rate of the 
Devetak-Winter protocol can be made positive. Using similar
filtration methods we show enhancement of `one way distillable
key rate' for two more families of states considered by Chi
\etal~\citep{DJJTS07}. In all the cases the enhancement is
dramatic and the one way distillable key rate goes from a
negative to a significant positive  value making the
transformed states explicitly usable for QKD. Since we
quantitatively calculate the key rate, the added advantage
is that the amount of key available can also be ascertained.

This paper is organized as follows: In
Section~\ref{background}, we recall the results that provide
background for our work, which includes the definition of private
states and distillable key rate given in
Section~\ref{private-states}, privacy squeezing and one way
distillable key  rate given in Section~\ref{privacy-squeezing}
and a description of local filtration process given in
Section~\ref{local-filters}. In Section~\ref{results} we
describe our main results where we show how the filtration
process can enhance the one way distillable key rate. Some
concluding remarks are presented in
Section~\ref{concluding}. 
\section{Background}
\label{background} 
In this section we briefly recapitulate
the results that our work builds upon. In this context
private states and their role in QKD protocols,
classical-classical-quantum (ccq) states,  the one way 
distillable key rate and local filters that we will use in the
next section for enhancement of the relevant key
rates of certain bound entangled states are discussed.
\subsection{Private states and distillable key}
\label{private-states}
Consider a situation where Alice and Bob want to carry out
QKD; Alice has systems $A$ and $A'$ of dimension  $d_{A}$
and $d_{A'}$ respectively, while Bob has systems $B$ and
$B'$ of dimension $d_{B}$ and $d_{B'}$, respectively.  The
key is contributed by the systems $A$ and $B$, therefore
$d_{A}=d_{B}=d$.  Let a state $\rho_{ABA'B'}\in
B(\hcal^d\ot\hcal^d\ot\hcal^{d_{A'}}\ot\hcal^{d_{B'}})$ be
shared between Alice and Bob and the eavesdropper has the
standard purifying system $E$. The purification
${|\psi_{\rho}}\>_{ABA'B'E}$ of the state $\rho_{ABA'B'}$ is
called ${\bf secure}$ if upon measurement on systems $A$ and
$B$ in the basis $\{|ij\>_{AB}\}_{i,j=0}^{d-1}$, followed by
tracing over the $A'B'$ subsystem,  the joint state of the
systems $A$, $B$ and $E$  takes the form of a
`classical-classical-quantum'  (ccq) state~:
\begin{equation} 
\rho^{\rm ccq}=\sum_{i,j=0}^{d-1} p_{ij}|i j\>
\<i j|_{AB}\otimes \rho_E.  
\label{eq:ccq-key}
\end{equation} 
The state $\rho_{ABA'B'}$ whose purification is  secure is
called  a {\bf private state} or a {\bf pdit} and $\rho^{\rm
ccq}$ is called the ccq-state corresponding to the secure
state.

The private states are related to maximally entangled states
of the system $AB$, and are  obtained by applying 
twisting operations  on the tensor product of a maximally
entangled state of $AB$ system and some arbitrary state
$\sigma_{A'B'}$ of the system $A'B'$ in the following
way:
\begin{equation} 
\gamma_{ABA'B'} = \frac{1}{d} \sum_{i,j=0}^{d-1} |i i\>\<j 
j|_{AB}\otimes U^{ii}\sigma_{A'B'} {U^{jj}}^{\dagger}.
\label{eq:pdit} 
\end{equation} 
The operators $U^{ii}$ are  arbitrary unitary
transformations on the $A'B'$ system.  Twisting does not
change the security related properties of a state as the
corresponding ccq-state is invariant under the twisting
operation. Therefore, the private states are as efficient
for QKD as maximally entangled states and have the same
amount of key. In fact given a private state we should
always be able to find a twisting operation so that we
recover the maximally entangled state of the $AB$ system by
twisting~\cite{HPHH,HHHO05a}.

Since, private states are analogous to maximally entangled
states in entanglement theory, similar to the definition of
measure of distillable entanglement, one can define maximal
distillable pdits. 
For any given state
$\rho_{AB}\in B(\hcal_A\otimes \hcal_B)$ let us consider a
sequence of $LOCC$ operations $\{P_n\}$ such that
$P_n(\rho_{AB}^{\ot n})=\phi_n$, where $\phi_n \in
B(\hcal^{(n)}_A\ot\hcal^{(n)}_B)$.  A set of operations
${\cal P} = \{P_n: n \in \mathbb{N}\}$ is called a pdit
distillation protocol of state $\rho_{AB}$ if  there is a
pdit $\gamma_{d_n}$  whose  key  part is of dimension
$d_n\times d_n$, satisfying 
\begin{equation}
\lim_{n\rightarrow \infty} ||\phi_n-\gamma_{d_n}|| = 0,
\end{equation} 
For a given distillation protocol $\cal P$, the key  rate is
given by
\begin{equation} 
{\cal R}({\cal P})=\limsup_{n\rightarrow
\infty} \frac{\log d_n}{n} 
\end{equation}
The {\bf distillable key rate} of state $\rho_{AB}$ is given
by maximizing over all possible protocols.
\begin{equation}
K_D(\rho_{AB})=\sup_{\cal P}{\cal R}(\cal P).
\label{kd}
\end{equation}
If one can transform by LOCC, such as the
recurrence protocol~\cite{dist-2}, sufficiently many
copies of a state $\rho_{ABA'B'}$ into a state close enough
to a private state in trace norm, then
$K_D(\rho_{ABA'B'})>0$.

It can be shown that there are PPT states for which the key
rate $K_D$ is positive~\cite{Curty04:key-ent,HHH98}.
In any case since separable states by construction do not
have any distillable key rate and PPT states are always
bound entangled, we can conclude that these states are bound
entangled states with non-zero distillable key rate. We will
consider a few of these states in the next section and the detailed
mathematical proofs are available in ~\cite{HHHO05a, DJJTS07}.
It is not always straightforward to calculate $K_D$ for a given
quantum state as an optimization over distillation protocols
in involved.
\subsection{One way distillable key rate} 
\label{privacy-squeezing}
In the special case when $d=2$, the private state is called a
{\bf pbit}, and we will restrict to this case in the rest of
this  paper.
In this scenario, consider a state $\rho_{ABA'B'}\in
B(\hcal^2\ot\hcal^2\ot\hcal ^{d_{A'}}\ot\hcal^{d_{B'}})$.
Expanding this state in the computational basis of the
qubits $A$ and $B$ namely, $\vert ij\rangle,$ with
$i,j\in\{0,1\}$:
\begin{equation} 
\rho_{ABA'B'}= \left[
\begin{array}{cccc} \sigma^{0000} & \sigma^{0001} & \sigma^{0010} &
\sigma^{0011} \\ \sigma^{0100} & \sigma^{0101} &
\sigma^{0110} & \sigma^{0111} \\
\sigma^{1000} & \sigma^{1001} & \sigma^{1010} &
\sigma^{1011} \\ \sigma^{1100} &
\sigma^{1101} & \sigma^{1110} & \sigma^{1111} 
\end{array} \right],
\label{aabb-state}
\end{equation}
The term
$|00\>\<11|_{AB}\otimes\sigma^{0011}$(upper right hand
corner in the above equation), with
$\sigma^{0011}$ being a density operator for the
$A'B'$ sub-system  plays an important role. If the state
$\rho_{ABA'B'}$ was such that
the $AB$ sub-system is in a maximally entangled state
$\frac{1}{\sqrt{2}}\left(\vert 00 \rangle + \vert 11
\rangle\right)$ then this term and its hermitian conjugate
namely the term $\vert 11\rangle\langle
00\vert_{AB}\otimes\sigma^{1100}_{A'B'}$ are the only
non-zero off diagonal terms in the expansion and the trace
norm $||\sigma^{1100}|| = ||\sigma^{0011}||=\frac{1}{2}$.
For any private state, by an appropriate twisting operation,
$||\sigma^{1100}||$ can always be brought to $\frac{1}{2}$.
For an arbitrary state $\rho_{ABA'B'}$, in order to determine its ability
to carry out QKD, we carry out an operation called privacy
squeezing where we maximize $||\sigma^{1100}||$ by applying
twisting operations.  The state $\rho^{\rm PS}$ that we
obtain after such a maximization is called  privacy squeezed
state corresponding to $\rho_{AA'BB'}$.
It has been shown by a more  detailed analysis~\cite{DW}
that
if $\|\sigma^{0000}\|=\|\sigma^{0011}\|=\|\sigma^{1111}\|$
and $\|\sigma^{0101}\|<\|\sigma^{0011}\|$,
$\|\sigma^{1010}\|<\|\sigma^{0011}\|$, then
$K_D(\rho_{ABA'B'})>0$.

For a specific form of $\rho_{ABA'B'} 
\in B(\hcal^2 \otimes \hcal^2 \otimes \hcal^2 \otimes \hcal^{2})$ 
\begin{eqnarray}
\rho_{ABA'B'}&=&\ket{\phi^+}\bra{\phi^+} \otimes \sigma_0
+\ket{\phi^-}\bra{\phi^-} \otimes \sigma_1\nonumber\\
&&+\ket{\psi^+}\bra{\psi^+} \otimes \sigma_2
+\ket{\psi^-}\bra{\psi^-} \otimes \sigma_3,
\label{eq:Belldiag_rho}
\end{eqnarray}
where  $\ket{\phi^{\pm}}$ and $\ket{\psi^{\pm}}$ are Bell
states in $\hcal^2\otimes \hcal^2$.
Then if $\|\sigma_0-\sigma_1\|> \frac{1}{2}$ and
$\T(\sigma_0\sigma_1)=0$,
then $K_{D}(\rho_{ABA'B'})>0$.

One-way key distillation protocol expresses the key distillation
rate for the ccq-state in terms of the Devetak-Winter
function $K^{DW}_{D}$~\cite{DW}.  Devetak-Winter function or
the one way distillable key rate, $K^{DW}_{D}$ is the
difference between the mutual informations of the Alice-Bob 
subsystem and the Alice-Eve subsystem: 
\begin{equation}
K^{DW}_{D}=I(A:B)-I(A:E)  
\end{equation}
where the mutual information between $A$ and $B$ is given as
$I(A:B)= S(A)+S(B)-S(AB)$ and between  $A$ and $E$ is
given as $I(A:E)=S(A)+S(E)-S(AE)$, with $S(X)$ being the
Von-Neumann entropy of the sub-system $X$.

For a state $\rho_{ABA'B'}$ of the form given in
Equation~(\ref{aabb-state}), let us 
define the following parameters
\begin{eqnarray}
x&=&(\|\sigma^{0000}\|+\|\sigma^{1111}\|)/2+\|\sigma^{0011}\|,\nonumber\\
y&=&(\|\sigma^{0000}\|+\|\sigma^{1111}\|)/2-\|\sigma^{0011}\|,\nonumber\\
z&=&(\|\sigma^{0101}\|+\|\sigma^{1010}\|)/2+\|\sigma^{0110}\|,\nonumber\\
w&=&(\|\sigma^{0101}\|+\|\sigma^{1010}\|)/2-\|\sigma^{0110}\|.
\label{eq:KDW00}
\end{eqnarray}
In terms of these parameters define
\begin{equation}
S(E)=-x\log_2x-y\log_2y-z\log_2z-w\log_2w,
\label{eq:SEabcd}
\end{equation}
The analysis given in~\cite{DW} proves that 
\begin{equation}
K_{D}^{DW}([\rho^{PS}]^{ccq}_{ABE'})= 1-S(E)
\label{kdw-formula}
\end{equation}
This is the key formula that we will use to compute 
the one way distillable key rates for various states in the
next section.

For any bipartite state $\rho_{ABA'B'} \in
{B}({\hcal}^2\ot{\hcal}^2 \ot {\hcal }^d\ot{\hcal}^{d'})$,
it can be shown that 
 \begin{equation} K_{D}(\rho_{ABA'B'}) \geq
K^{DW}_{D}([\rho^{PS}]^{ccq}_{ABE'})
\label{kdw>kd} 
\end{equation}
 where
$[\rho^{PS}]^{\rm ccq}_{ABE'}$ is the  ccq-state
corresponding to the privacy squeezed state of
$\rho_{ABA'B'}$.  Combining Equation~(\ref{eq:SEabcd})  with
Equation~(\ref{kdw>kd}) will allow us to prove the positive
distillable key rates of several bound entangled states.

\subsection{Local Filters}
\label{local-filters}
In a local filtering process that we consider,
Alice and Bob perform local POVMs on the state $\rho_{AA'BB'} 
\in B({\hcal }^2\ot{\hcal}^2 \ot {\hcal }^2\ot{\hcal}^{2})$
defined by the operators $L_A$ and $L_B$. They retain the
cases when both of them get positive outcome for the
operator $L_A$ and $L_B$ and discard the other cases.
The state after such an operation changes as follows:
\begin{equation}
\rho \rightarrow (L\rho
L^{\dag})/(\tr{L\rho L^{\dag}}), \quad
L=L_A\otimes L_B
\label{filter}
\end{equation}
In our case $L_A$ and $L_B$ are $4\times 4$ diagonal  matrices
leading to 
\begin{eqnarray}
L=\left[%
\begin{array}{cccc}
 a & 0 & 0 & 0 \\
 0 & b & 0 & 0 \\
 0 & 0 & c & 0 \\
 0 & 0 & 0 & d
\end{array}
\right]_{A}
\otimes
\left[%
\begin{array}{cccc}
 r & 0 & 0 & 0 \\
 0 & s & 0 & 0 \\
 0 & 0 & t & 0 \\
 0 & 0 & 0 & u
\end{array}
\right]_{B}
\label{filter-1}
\end{eqnarray}
The $16\times 16$ matrix $L$ can also be written as 
\begin{equation}
L=\left[%
\begin{array}{cccc}
  L_{1} & 0 & 0 & 0\\
  0 & L_{2} & 0 & 0 \\
  0 & 0 & L_{3} & 0 \\
  0 & 0 & 0 & L_{4} \\
\end{array}%
\right]
\label{filter-2}
\end{equation}
where $L_1$, $L_2$, $L_3$ and $L_4$ all are $4\times 4$
diagonal matrices
determined in terms of $L_A$ and $L_B$  and
we have
\begin{eqnarray}
L_{1}=\left[%
\begin{array}{cccc}
 a r & 0 & 0 & 0 \\
 0 & a s & 0 & 0 \\
 0 & 0 & a t & 0 \\
 0 & 0 & 0 & a u
\end{array}
\right],
&
L_{2}=\left[
\begin{array}{cccc}
 b r & 0 & 0 & 0 \\
 0 & b s & 0 & 0 \\
 0 & 0 & b t & 0 \\
 0 & 0 & 0 & b u
\end{array}
\right], \nonumber \\
L_{3}=\left[
\begin{array}{cccc}
 c r & 0 & 0 & 0 \\
 0 & c s & 0 & 0 \\
 0 & 0 & c t & 0 \\
 0 & 0 & 0 & c u
\end{array}
\right], &
L_{4}=\left[
\begin{array}{cccc}
 d r & 0 & 0 & 0 \\
 0 & d s & 0 & 0 \\
 0 & 0 & d t & 0 \\
 0 & 0 & 0 & d u
\end{array}
\right].
\label{filter-3}
\end{eqnarray}
The parameters
$\{a,b,c,d,r,s,t,u\}\in(0,1]$ define the local filter. The
filtration is typically considered as an  operator on a
large ensemble of identically prepared states
$\rho_{AA'BB'}$ and lead to smaller ensemble of filtered
states. The success probability $P$ of a filter
is defined as the probability with which both the
measurement give the positive result together and the size
of the ensemble after filtration is $P$ times the size of
the original ensemble.
\section{Enhancement of distillable key of bound entangled
states by local filtration}
\label{results}
We are now ready to consider the possibility of enhancing
the distillable key rate from bound entangled states by a
local filtration process.  The starting states that we
consider are bound entangled because they are PPT and the
local filters cannot change states from PPT to NPT and hence
the filtered states are also bound entangled.  We then
analyze the distillable key rate from the new and old states
by specifically computing the one way distillable key rate
defined in Devetak-Winter protocol by a procedure described
in Section~\ref{background}. In every case the one
distillable key rate goes from a negative to a positive
value.
\subsection{Example 1}
\begin{figure}[t]
\includegraphics[scale=1]{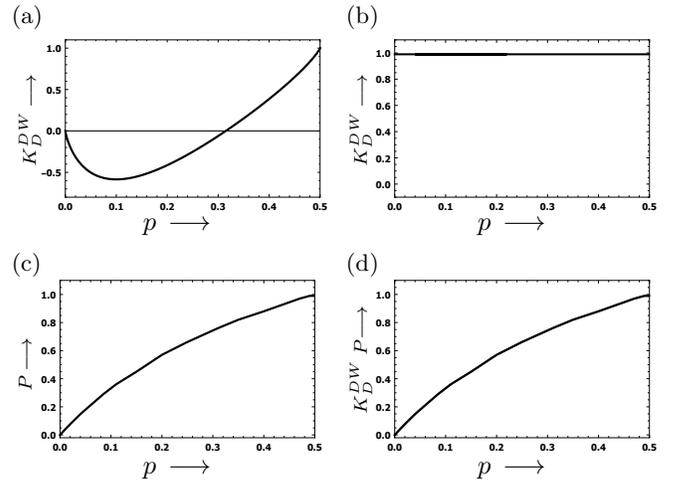}
\caption{Improvement in the  value of $K_{D}^{DW}$ after
local filtration operation on the one parameter family of
quantum states
$\rho^{(1)}_p$. (a) Value of the one way distillable key rate
$K_{D}^{DW}$ for the family of states $\rho^{(1)}_p$ as a function 
of $p$. (b)
The value of  $K_{D}^{DW}$ after optimal local filtration operation
on $\rho^{(1)}_p$.
(c) The probability of success $P$ of the optimal local filter.
(d) The effective one way distillable key  rate after
filtration obtained by multiplying the value of 
$K_{D}^{DW}$ with the success probability of the optimal
filter. 
For all plots the $x$-axis is the parameter $p \in [0,1/2]$.
The enhancement of the one way distillable 
key rate can be seen by comparing the graphs shown in (a) and (d).
\label{fig4_1}}
\end{figure}
Consider a one parameter family of states $\rho^{(1)}_p 
\in B({\hcal }^2\ot{\hcal}^2 \ot {\hcal }^2\ot{\hcal}^{2})$
parameterized by  a real parameter with $p \in [0,1/2]$ 
define as
\begin{equation}
\rho^{(1)}_p= 
\frac{1}{1+2p}
\left[%
\begin{array}{cccc}
2p \tau_1 &0&0&2p \tau_1\\
0& \left(\frac{1}{2}-p\right)\tau_2&0&0 \\
0&0&\left(\frac{1}{2}-p\right)\tau_2& 0\\
2p \tau_1 &0&0&2p \tau_1\\
\end{array}%
\right]
\label{the_new_example}
\end{equation}
with 
 matrices $\tau_1$ and
$\tau_2$ given as 
\begin{equation}
\tau_1 =
\left[
\begin{array}{rrrr}
 \frac{1}{6} & 0 & 0 & 0 \\
 0 & \frac{1}{3} & -\frac{1}{6} & 0 \\
 0 & -\frac{1}{6} & \frac{1}{3} & 0 \\
 0 & 0 & 0 & \frac{1}{6} \\
\end{array}
\right] \, {\rm and}\quad
\tau_2 =
\left[
\begin{array}{rrrr}
 \frac{1}{3} & 0 & 0 & 0 \\
 0 & \frac{1}{6} & \frac{1}{6} & 0 \\
 0 & \frac{1}{6} & \frac{1}{6} & 0 \\
 0 & 0 & 0 & \frac{1}{3} \\
\end{array}
\right].
\label{tau12}
\end{equation}
This
family of state is PPT and hence any entanglement if
present, has to be bound entanglement.

Since this  family of states belongs to 
$B({\hcal }^2\ot{\hcal}^2 \ot {\hcal }^2\ot{\hcal}^{2})$
as per
Equation~(\ref{kdw>kd}), $K_D$ is always greater than
$K^{DW}$ and therefore, $K^{DW}_D$ provides us with a  lower
bound for the distillable key rate. 
We calculate the value of $K_D^{DW}$ for these states by
exploiting the fact that the structure of the family of
states is the same as
that given in Equation~(\ref{aabb-state}),  for which we can
calculate $K_D^{DW}$. Comparison with
Equation~(\ref{eq:KDW00}) reveals that for this family of
 states we have  $x=4\left| \frac{p}{1+2p} \right| $,
$y=0$, $z= \frac{1}{2}\left|
\frac{-1+2p}{1+2p}\right|$ and $w= \frac{1}{2}\left|
\frac{-1+2p}{1+2p}\right|$.
Therefore, using the formula~(\ref{eq:SEabcd}) and
Equation~(\ref{kdw-formula}) we can calculate
$K_{D}^{DW}([\rho^{(1)}_p~^{PS}]^{ccq}_{ABE'})$.  The graph
of $K_{D}^{DW}([\rho^{(1)}_p~^{PS}]^{ccq}_{ABE'})$ as a
function of $p$ is plotted in Figure~\ref{fig4_1}(a).  The
value of $K_D^{DW}$ is negative upto a certain  values of
$p$ and then crosses over to a positive value at approximately 
$p=0.31$.

Now, we propose local filtration operation on the family of
states $\rho^{(1)}_p$ with a view to raise the value of
the corresponding $K_{D}^{DW}([\rho^{(1)}_p~^{PS}]^{ccq}_{ABE'})$.
Consider the application of the local filter described in
Equations~(\ref{filter}) 
on the state $\rho^{(1)}_p$.  The
state after filtration is  given by 
\begin{equation}
\rho^{(1)}_p \rightarrow \frac{L\rho^{(1)}_p
L^{\dag}}{\tr({L\rho^{(1)}_p L^{\dag})}} = \rho^{(1)'}_p.
\label{filter1} 
\end{equation}
Exploiting the structure of the filtration operation 
detailed in Equations~(\ref{filter-1}), (\ref{filter-2}) \&
(\ref{filter-3})
we can write $\rho^{(1)'}_p$ in explicit matrix form as:
\begin{eqnarray}
&&\rho^{(1)'}_p=
\frac{1}{(1+2p)M}\times
\nonumber \\
&&\!\!\left[%
\begin{array}{cccc}
\!L_{1}(2p{\tau_1})L^{\dag}_{1} & 0 & 0 &
\!L_{1}(2p{\tau_1})L^{\dag}_{4} \\
0 & \!L_{2}(\frac{1}{2}-p)\tau_2 L^{\dag}_{2} & 0 & 0 \\
0 & 0 & \!L_{3}(\frac{1}{2}-p)\tau_2 L^{\dag}_{3} & 0 \\
\!L_{4}(2p{\tau_1})L^{\dag}_{1} & 0 & 0 &
\!L_{4}(2p{\tau_1})L^{\dag}_{4} \\
\end{array}%
\right].\nonumber\\
\end{eqnarray}
Where $M$=$\tr{(L\rho_{p}^{(1)} L^{\dag})}$.
For calculating the one way distillable key rate we use the following 
Equation~(\ref{eq:KDW00}) to ascertain the values of the
parameters $x$, $y$, $z$ and
$w$ for this state as
\begin{eqnarray}
x&=&(\| L_{1}2p({\tau_1})L^{\dag}_{1}/(1+2p)M \|\nonumber\\
&+&\|L_{4}2p({\tau_1})L^{\dag}_{4}/(1+2p)M\|)/2\nonumber\\
&+&\|L_{1}2p({\tau_1})L^{\dag}_{4}/(1+2p)M\|,\nonumber\\
y&=&(\|
L_{1}2p({\tau_1})L^{\dag}_{1}/(1+2p)M\|\nonumber\\
&+&\|L_{4}2p({\tau_1})L^{\dag}_{4}/(1+2p)M\|)/2\nonumber\\
&-&\|L_{1}2p({\tau_1})L^{\dag}_{4}/(1+2p)M\|,\\
z&=&( \|L_{2}(\frac{1}{2}-p)\tau_2 L^{\dag}_{2}/(1+2p)M\|\nonumber\\
&+&\|L_{3}(\frac{1}{2}-p)\tau_2 L^{\dag}_{3}/(1+2p)M\|)/2,\nonumber\\
w&=&( \|L_{2}(\frac{1}{2}-p)\tau_2 L^{\dag}_{2}/(1+2p)M\|\nonumber\\
&+&\|L_{3}(\frac{1}{2}-p)\tau_2 L^{\dag}_{3}/(1+2p)M\|)/2.\nonumber
\end{eqnarray}
The value of $K_D^{DW}([\rho^{(1)}_p~^{PS}]^{ccq}_{ABE'})$ can be
again calculated following Equation~(\ref{eq:SEabcd}) and
(\ref{kdw-formula}), as was done before filtering.

A given filter has a success probability $P$ which is
defined as the fraction of cases the filter gives a positive
answer. In real terms, for the purposes of using an ensemble
of states for QKD we must multiply the key rate with the
success probability to get the effective key rate. This is
particularly important if  we want to make any comparison
with the key rate before filtration. A filter may give us a
very high key rate, however, if its probability of success
is small the effective key rate may in fact be small.  For a
given value of $p$ we numerically optimize the product
$K_D^{DW} P$ to obtain the optimal filter.
It turns out that for this family of states the parameters
of the optimal filters are 
\begin{eqnarray}
&a=d=r=s=t=u=1&\nonumber \\
&b=c=( \text{small $p$ dependent  value}).&
\label{opt_filter}
\end{eqnarray}

The results are shown in Figure~\ref{fig4_1}.  In
Figure~\ref{fig4_1}(a) the value of one way distillable key
rate $K_{D}^{DW}$ for the family of states $\rho^{(1)}_p$ is
plotted as a function of $p\in[0,1/2]$ before filtration.
This value is clearly negative for $p$ values upto a
certain value and becomes positive only after $p$
approximately crosses $0.31$. In Figure~\ref{fig4_1}(b) The
value of  $K_{D}^{DW}$ is plotted as a function $p$ after
optimal local filtration operation. It is noteworthy that we
are always able to find a filter such that the value of the
one way  distillable key rate is equal to its maximum
possible value $1$.  In Figure~\ref{fig4_1}(c) we plot the
success probability $P$ of the optimal local filter as a function
of $p$ which clearly is small for small values of $p$ but
quickly becomes significant as the value of $p$
increases and approaches $1$ as $p$ approaches $0.5$.  In
Figure~\ref{fig4_1}(d) we plot the  effective one way 
distillable key rate obtained by multiplying  $K_{D}^{DW}$ and
$P$ for the optimal filter as a function $p$. It is clear
form the graphs that the filtration process is able to make the
one way distillable key  rate positive for all values $p$.
We must compare the values in Figure~\ref{fig4_1}(a) with
values in Figure~\ref{fig4_1}(d) to access the enhancement
achieved through filtration. The comparison clearly shows
that every member of the family of states $\rho_p^{(1)}$
after filtration has a positive value of $K_{D}^{DW}$.  This
implies that this family of PPT states have non-zero one way
distillable key rate after filtration.  In other words,
$\rho^{(1)}_p$ is a family of  bound entangled
states with a nonzero value of $K_D^{DW}$.  Since for this
class of states $K_D > K_D^{DW}$ they thus have a nonzero
distillable key rate.  Any PPT entangled state remains PPT
entangled under a local filtration operation, this implies
that all states in the original family(before filtration)
are PPT entangled states which can be employed for QKD after
filtration.

As it turns out this family of states is closely related to
the states considered and analyzed  by K. Horodecki
\etal~\cite{HHHO05a}.  They consider states $\rho_{(p,d,k)}$
which for $d=2,k=1$ are defined as 
\begin{multline}
\rho_{(p,2,1)}= \\ \begin{bmatrix}
\frac{p}{2} (\tau_1+\tau_2) &0&0&\frac{p}{2}(\tau_1-\tau_2) \\
0& \left(\frac{1}{2}-p\right)\tau_2&0&0 \\
0&0&\left(\frac{1}{2}-p\right)\tau_2& 0\\
\frac{p}{2}(\tau_1-\tau_2) &0&0&\frac{p}{2} (\tau_1+\tau_2)  \\
\end{bmatrix},
\label{row-key-new}
\end{multline}
Consider a projection operator $A$ such that
\begin{equation}
A =\frac{1}{2}\left[
\begin{array}{cccc}
I & 0 & 0 & I\\
0 & I & 0 & 0 \\
0 & 0 & I & 0 \\
I & 0 & 0 & I \\
\end{array}
\right].
\end{equation}
Where  $I$  is an identity operator on a $2\ot2$ space.
The families of states $\rho^{(1)}_p$ and
$\rho_{(p,2,1)}$ are connected with each other
via the operator $A$ as follows:
\begin{equation}
\rho^{(1)}_p= \frac{A\rho_{(p,2,1)}
A^{\dag}}{Tr({A\rho_{(p,2,1)} A^{\dag})}}.
\end{equation}
\subsection{Example 2}
\begin{figure}[t]
\includegraphics[scale=1]{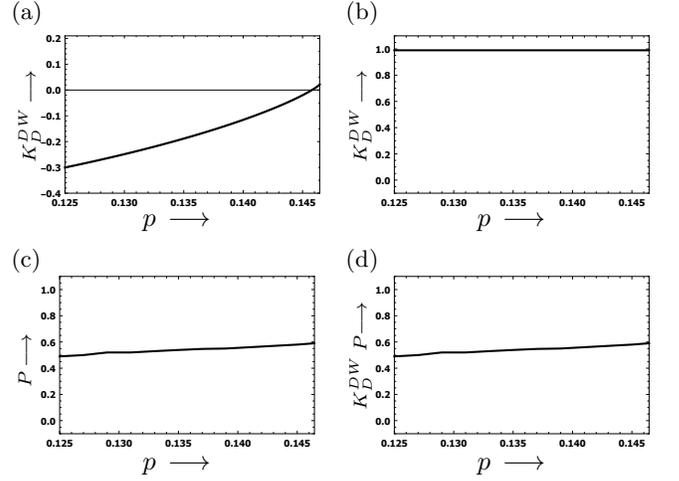}
\caption{Results for the the family of states
$\rho^{(2)}_p$.  Various quantities are plotted as a
function of the parameter $p$ in the range
$(\frac{1}{8},\frac{1}{4+2\sqrt{2}})$.  (a) Displays
$K^{DW}_D$ for the states $\rho^{(2)}_p$  before
filtration.  (b) Displays $K^{DW}_D$ for $\rho^{(2)'}_p$
which is the state  after optimal filtration.  (c) Displays
the success probability $P$ of the optimal filter.
(d) Displays the product  $K^{DW}_D P$ which represents the
effective key rate after filtration.  Comparison of the plot
(a) and (d) clearly shows that an  enhancement of key rate
has been achieved by filtration.
\label{fig2}} 
\end{figure}  
Next we consider examples of families of states  considered
by Chi
\etal~\cite{DJJTS07}.
The family of states $\rho^{(2)}_p  \in
B({\hcal }^2\ot{\hcal}^2 \ot {\hcal }^2\ot{\hcal}^{2})$
is defined as follows:
\begin{equation}
\rho^{(2)}_p
=\frac{1}{2}\left[%
\begin{array}{cccc}
\sigma_0+\sigma_1 & 0 & 0 & \sigma_0-\sigma_1 \\
0 & \sigma_2+\sigma_3 & \sigma_2-\sigma_3 & 0 \\
0 & \sigma_2-\sigma_3 & \sigma_2+\sigma_3 & 0 \\
\sigma_0-\sigma_1 & 0 & 0 & \sigma_0+\sigma_1 \\
\end{array}%
\right].
\end{equation}
with
\begin{equation}
\begin{array}{c}
\sigma_0=\frac{1}{2}\left[%
\begin{array}{cccc}
p & 0 & 0 & p \\
0 & 2 p & 0 & 0 \\
0 & 0 & 0 & 0 \\
p & 0 & 0 & p \\
\end{array}%
\right], \,\,
\sigma_1 =
\frac{1}{2}\left[
\begin{array}{cccc}
p & 0 & 0 & -p \\
0 & 0 & 0 & 0 \\
0 & 0 & 2 p & 0 \\
-p & 0 & 0 & p 
\end{array}
\right],\\
\sigma_2= \frac{1}{2}\left[
\begin{array}{cccc}
1-4p-2\sqrt{2}p   & 0 & 0 & 0 \\
  0 & (\sqrt{2}+1)p & p & 0 \\
  0 & p & (\sqrt{2}-1)p& 0 \\
  0 & 0 & 0 & 0 \\
\end{array}%
\right],\\
\sigma_3
=\frac{1}{2}\left[
\begin{array}{cccc}
1-4p-2\sqrt{2}p & 0 & 0 & 0 \\
0 & (\sqrt{2}-1)p & -p & 0 \\
0 & -p & (\sqrt{2}+1)p & 0 \\
0 & 0 & 0 & 0 \\
\end{array}%
\right]
\end{array}
\label{sigma_matrices}
\end{equation}
This family of states is determined by single real parameter
$p$ in the range
$(\frac{1}{8},\frac{1}{4+2\sqrt{2}})$. It is
straightforward to see that the states in the family are PPT
as $(\rho^{(2)}_p)^\Gamma=\rho^{(2)}_p$ and therefore, if
they have entanglement it has to be bound
entanglement~\cite{HHH98}.  Since the family of states have
a form given in Equation~(\ref{eq:Belldiag_rho}) and here we
have $\|\sigma_0+\sigma_1\| = 4p$ and $\|\sigma_0-\sigma_1\|
= 4p > 1/2$, we have $K_D(\rho^{(2)}_p)>0$~\cite{DJJTS07}.

In order to calculate the one way distillable key rate for
this family of states, we first identify the  parameters 
$x$, $y$, $z$, and $w$ as per
Equation~(\ref{eq:KDW00}) which turn out to be
\begin{eqnarray}
&&x=4p,\quad y=0, \nonumber \\
&& z=\frac{1-4p+2\sqrt{2}p}{2}
\,{\rm and} \quad w=\frac{1-4p-2\sqrt{2}}{2},
\end{eqnarray}
Using Equation~(\ref{eq:SEabcd})\&(\ref{kdw-formula}) we
obtain the value of 
\begin{multline}
K_D^{DW}([\rho^{(1)}_p~^{PS}]^{ccq}_{ABE'})=1+4p\log_24p\\
+\left(\frac{1-4p-2\sqrt{2}}{2}\right)\log_2\left(\frac{1-4p-2\sqrt{2}}{2}\right)\\
+\left(\frac{1-4p+2\sqrt{2}p}{2}\right)\log_2\left(\frac{1-4p+2\sqrt{2}p}{2}\right)
\label{eq:ex36}
\end{multline}
Here $[\rho^{(2)}_{p}~^{PS}]^{ccq}_{ABE'}$ is the ccq state for the
privacy squeezed state of $\rho^{(2)}_p$.
For this class of
states the values of $K_D^{DW}$ is negative for most value 
of $p$, therefore, although $K_D>0$ utility of the states
for QKD is not clear. The exact plot of $K_D^{DW}$ as
function of $p$ is shown in Figure~\ref{fig2}(a).

Next we apply filtration operation on $\rho^{(2)}_p$ as
described in Equation~(\ref{filter}), 
\begin{equation}
\rho^{(2)}_p \rightarrow \frac{L\rho^{(2)}_p
L^{\dag}}{\tr{L\rho^{(2)}_p L^{\dag}}} = \rho^{(2)'}_p
\end{equation}
By using the form of filtration matrices given in
Equation~(\ref{filter-1})\&(\ref{filter-2})
the density operator after filtration operation can be
written as:
\begin{widetext}
\begin{equation}
\rho^{(2)'}_p=\frac{1}{2M}\left[%
\begin{array}{cccc}
L_{1}(\sigma_0+\sigma_1)L^{\dag}_{1} & 0 & 0 &
L_{1}(\sigma_0-\sigma_1)L^{\dag}_{4} \\ 0 &
L_{2}(\sigma_2+\sigma_3)L^{\dag}_{2} &
L_{2}(\sigma_2-\sigma_3)L^{\dag}_{3} & 0 \\ 0 &
L_{3}(\sigma_2-\sigma_3)L^{\dag}_{2} &
L_{3}(\sigma_2+\sigma_3)L^{\dag}_{3} & 0 \\
L_{4}(\sigma_0-\sigma_1)L^{\dag}_{1} & 0 & 0 &
L_{4}(\sigma_0+\sigma_1)L^{\dag}_{4} \\ \end{array}%
\right].\\ \end{equation}\end{widetext}
Each element in the above expression is a $4\times 4$ matrix and
can be computed by using Equation~(\ref{filter-3}),
here $M$=$\tr{(L\rho_{p}^{(2)} L^{\dag})}$.
In order to calculate the 
value of $K_{D}^{DW}([\rho^{(2)'}_p~^{PS}]^{ccq}_{ABE'})$,
we first map the parameters $x$,$y$,$z$ and $w$ as per
Equation~(\ref{eq:KDW00}) which turn
out to be 
\begin{eqnarray}
x&=&(\|L_{1}(\sigma_0+\sigma_1)L^{\dag}_{1}/2M\|+
\|L_{4}(\sigma_0+\sigma_1)L^{\dag}_{4}/2M\|)/2\nonumber\\
&+&\|L_{1}(\sigma_0-\sigma_1)L^{\dag}_{4}/2M\|,\nonumber\\
y&=&(\|L_{1}(\sigma_0+\sigma_1)L^{\dag}_{1}/2M\|+
\|L_{4}(\sigma_0+\sigma_1)L^{\dag}_{4}/2M\|)/2\nonumber\\
&-&\|L_{1}(\sigma_0-\sigma_1)L^{\dag}_{4}/2M\|,\nonumber\\
z&=&( \|L_{2}(\sigma_2+\sigma_3)L^{\dag}_{2}/2M\|
+\|L_{3}(\sigma_2+\sigma_3)L^{\dag}_{3}/2M\|)/2\nonumber\\
&+&\|L_{2}(\sigma_2-\sigma_3)L^{\dag}_{3}/2M\|,\nonumber\\
w&=&( \|L_{2}(\sigma_2+\sigma_3)L^{\dag}_{2}/2M\|
+\|L_{3}(\sigma_2+\sigma_3)L^{\dag}_{3}/2M\|)/2\nonumber\\
&-&\|L_{2}(\sigma_2-\sigma_3)L^{\dag}_{3}/2M\|.
\end{eqnarray} 
We now calculate the value of $
K_D^{DW}([\rho^{(2)'}_p~^{PS}]^{ccq}_{ABE'})$ by once again
using Equations~(\ref{eq:SEabcd})\&(\ref{kdw-formula}).

The filter is numerically optimized to maximize the
effective key rate which is the product of the 
one way distillable key rate 
$K_D^{DW}$ 
and the success probability of the filter $P$.
The structure of the filter turns out to be similar to what 
was obtained  for the previous example and is give in 
Equation~(\ref{opt_filter}). The results are displayed
in  
Figure~\ref{fig2}, where in 
Figure~\ref{fig2}(a)  we plot 
$K_D^{DW}$ before filtration, in Figure~\ref{fig2}(b) we
plot 
$K_D^{DW}$ after filtration, in Figure~\ref{fig2}(c) we
plot the success probability of the filter and in
Figure~\ref{fig2}(d) we plot the effective key rate which is
the quantity 
$K_D^{DW} P$, as functions of $p$. It is clear from a
comparison of plots 
Figure~\ref{fig2}(a) and Figure~\ref{fig2}(d) that for the
entire family of states the one way distillable key rate
turns positive from negative. 
By using Devetak-Winter lower bound of distillable key rate
for the ccq states we can therefore state that 
distillable key rate   $K_D(\rho^{(2)'}_p) \ge 
K_D^{DW}([\rho^{(2)'}_p~^{PS}]^{ccq}_{ABE'})>0$. This
establishes the usefulness of this family of states for QKD
and provides a quantitative estimate of the available key.
\subsection{Example 3} 
\begin{figure}[t]
\includegraphics[scale=1]{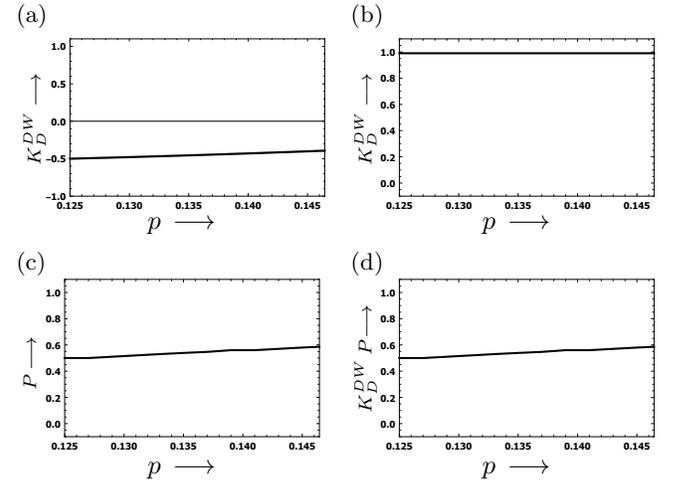}
\caption{Results for the family of states $\rho^{(3)}_p$.
The plots of $K_{D}^{DW}$ before and after filtration are
shown in parts (a) and (b), the success probability $P$ of the
optimal filter is shown in (c) while in (d) the effective
key rate is displayed in a way similar to
Figures~\ref{fig4_1} and~\ref{fig2}.  The one way key rate
change from completely negative to positive is clearly
evident.} \label{fig3} 
\end{figure} 

We consider another example from the paper of Chi 
\etal~\cite{DJJTS07} namely the family of states
$\rho^{(3)}_p$  given by the density
operator:
\begin{equation}
\rho^{(3)}_p=\frac{1}{2}\left[%
\begin{array}{cccc}
{\sigma_0+\sigma_1} & 0 & 0 &  {\sigma_0-\sigma_1} \\
0 & 2\sigma_2 & 0 & 0 \\
0 & 0 & 2\sigma_2 & 0 \\
{\sigma_0-\sigma_1} & 0 & 0 &  {\sigma_0+\sigma_1} \\
\end{array}%
\right].
\label{eq:ex40}
\end{equation}
The operators $\sigma_0$ and $\sigma_1$ are defined
in Equation~(\ref{sigma_matrices}) and
$\sigma_2=\frac{p}{\sqrt{2}}(|01\rangle\langle
01|+|10\rangle\langle 10|)+
(\frac{1}{4}-(1+\frac{1}{\sqrt{2}})p)(|00\rangle\langle
00|+|11\rangle\langle 11|)$. The states in this one
parameter family of states are well defined and are bound
entangled states in the range $\frac{1}{8} \leq
p \leq \frac{1}{4+2\sqrt{2}}$.  Again
as per~\cite{DJJTS07} for this family of states
$K_{D}(\rho^{(3)}_p)>0$. However by an explicit
calculation it was shown that 
$K_{D}^{DW}([\rho^{(3)}_{p}~^{PS}]^{ccq}_{ABE'})<0$ for
$\frac{1}{8} \leq p \leq \frac{1}{4+2\sqrt{2}}$.

As was done in  the earlier cases we 
apply filtration operation on $\rho^{(3)}_p$ as described
in Equation~(\ref{filter}) 
\begin{equation}
\rho^{(3)}_p \rightarrow \frac{L\rho^{(3)}_p
L^{\dag}}{\tr{L\rho^{(3)}_p L^{\dag}}} = \rho^{(3)'}_p
\end{equation} 
Again the explicit form of the density operator can be
written in terms of filter parameters explicitly as:
written as:
\begin{eqnarray}
&&\rho^{(3)'}_p=\frac{1}{2M}\times
\nonumber \\
&&\left[%
\begin{array}{cccc}
L_{1}(\sigma_0+\sigma_1)L^{\dag}_{1} & 0 & 0 &
L_{1}(\sigma_0-\sigma_1)L^{\dag}_{4} \\ 
0 & L_{2}(2\sigma_2)L^{\dag}_{2} & 0 & 0 \\ 
0 & 0 & L_{3}(2\sigma_2)L^{\dag}_{3} & 0 \\
L_{4}(\sigma_0-\sigma_1)L^{\dag}_{1} & 0 & 0 &
L_{4}(\sigma_0+\sigma_1)L^{\dag}_{4} \\ \end{array}%
\right].\nonumber \\ 
\end{eqnarray}
Each element in the above expression is a $4\times 4$ matrix
and
can be computed by using Equation~(\ref{filter-3}), here
$M$=$\tr{(L\rho_{p}^{(3)} L^{\dag})}$.
In order to calculate the 
value of $K_{D}^{DW}([\rho^{(3)'}_{p}~^{PS}]^{ccq}_{ABE'})$,
we first map the parameters $x$,$y$,$z$ and $w$ as per
Equation~(\ref{eq:KDW00}) which turn
out to be 
\begin{eqnarray}
x&=&(\|L_{1}(\sigma_0+\sigma_1)L^{\dag}_{1}/2M\|+
\|L_{4}(\sigma_0+\sigma_1)L^{\dag}_{4}/2M\|)/2\nonumber\\
&+&\|L_{1}(\sigma_0-\sigma_1)L^{\dag}_{4}/2M\|,\nonumber\\
y&=&(\|L_{1}(\sigma_0+\sigma_1)L^{\dag}_{1}/2M\|+
\|L_{4}(\sigma_0+\sigma_1)L^{\dag}_{4}/2M\|)/2\nonumber\\
&-&\|L_{1}(\sigma_0-\sigma_1)L^{\dag}_{4}/2M\|,\nonumber\\
z&=&( \|L_{2}(2\sigma_2)L^{\dag}_{2}/2M\|
+\|L_{3}(2\sigma_2)L^{\dag}_{3}/2M\|)/2\nonumber\\
w&=&( \|L_{2}(2\sigma_2)L^{\dag}_{2}/2M\|
+\|L_{3}(2\sigma_2)L^{\dag}_{3}/2M\|)/2.\nonumber\\
\end{eqnarray} 
The one way distillable key rate is calculated for the
numerically optimized filter which in this case again has
the structure given in Equation~(\ref{opt_filter}).  The results
are displayed in different parts of Figure~\ref{fig3}, where
$K_D^{DW}$ before and after filtration operation, the
success probability of the filter and the effective key rate
( $K_D^{DW} P$) are plotted as functions of $p$ in the
relevant range of the parameter $p$. A comparison of
Figure~\ref{fig3}(a) and Figure~\ref{fig3}(d) make it clear
that the one way  distillable key rate has turned from negative to
positive under filtration and we now have a filtered
ensemble ready for QKD.
\section{Concluding Remarks}
\label{concluding}
In this paper we have explored the role of local filtration
operations in enhancing the distillable key rate available
from bound entangled states for QKD. The route we took was
to calculate the Winter-Devetak function (also called
the one way distillable key rate) which provides a lower bound 
for the distillable key rate for the examples that we
considered. Three examples of families of bound entangled
states were analyzed from this point of view and in each
case a significant enhancement of the key rate was achieved
by the filtration process. The filter was optimized so as to
maximize the post filtration one way distillable key rate
multiplied by the success probability of the filter. The one
way distillation key rate in each case turned from negative
to positive and the effective key rate was significant. Our
results provide quantitative estimates of available key rates
for the family of states that we consider. 
\section*{Acknowledgement}
R.S. acknowledges financial supports from
DST/ICPS/QuST/Theme-2/2019/General Project number
{\sf Q-90}. Arvind acknowledges the financial
support from DST/ICPS/QuST/Theme-1/2019/General Project
number {\sf Q-68}.
%

\end{document}